\documentclass[11pt]{article}

\usepackage[latin1]{inputenc}
\usepackage{amsfonts}
\usepackage[dvips]{graphics}
\usepackage{graphicx}
\usepackage{cite}
\usepackage{stmaryrd}
\usepackage{amssymb}
\usepackage{amsmath}
\usepackage{fancyhdr}
\usepackage[body={6.0in, 8.2in},left=1.25in,right=1.25in,bottom=0.65in]{geometry}
\usepackage{setspace}                    
\usepackage{hyperref}

\def\eps{\epsilon}

\def\sd{\partial\hspace{-6pt}\slash\hspace{2pt}}
\def\nablas{\nabla\hspace{-6pt}\slash\hspace{2pt}}
\def\scS{\cS\hspace{-6pt}\slash\hspace{2pt}}
\def\xis{\xi\hspace{-6pt}\slash\hspace{2pt}}
\def\psis{\psi\hspace{-6pt}\slash\hspace{2pt}}
\def\epss{\eps\hspace{-5pt}\slash\hspace{2pt}}
\def\div{\partial \cdot}

\def\d{\partial}

\renewcommand{\exp}[1]{{\rm exp}(#1)}
\def\half{{\frac{1}{2}}}

\def\cD{{\cal D}}

\def\cF{{\cal F}}
\def\cG{{\cal G}}

\def\cL{{\cal L}}

\def\cQ{{\cal Q}}
\def\cR{{\cal R}}
\def\cS{{\cal S}}

\def\beq{\begin{equation}}
\def\eeq{\end{equation}}

\def\Slash#1{#1\hspace{-6pt}\slash\hspace{2pt}}

\author{\large{N.~Bouatta$^a$, G.~Comp\`{e}re$^a$\footnote{Research Fellow of the National Fund for
Scientific Research (Belgium)}\ \ and \ A.~Sagnotti$^b$
   \footnote{Corresponding author. Email: Augusto.Sagnotti@roma2.infn.it.}}\\ \\
    \small{\it{$^a$Physique th\'{e}orique et math\'{e}matique and}}\\
    \small{\it{International Solvay Institutes}} \\
 \small{\it{Universit\'{e} libre de Bruxelles}} \\
  \small{\it{Campus de la Plaine CP 231, B-1050 Bruxelles, Belgium}}\\\\
 \small{\it{$^b$Dipartimento di Fisica}}\\\small{\it{ Universit\`{a} di Roma "Tor Vergata"}}\\
 \small{\it{INFN, Sezione di Roma "Tor Vergata" }}\\
 \small{\it{Via della Ricerca Scientifica 1, 00133 Roma, Italy }}
 }

\title{\Huge{An Introduction to Free Higher-Spin Fields}}

\begin{document}

\begin{titlepage}

\maketitle

\begin{abstract}
In this article we begin by reviewing the (Fang-)Fronsdal
construction and the non-local geometric equations with
unconstrained gauge fields and parameters built by Francia and the
senior author from the higher-spin curvatures of de Wit and
Freedman. We then turn to the triplet structure of totally
symmetric tensors that emerges from free String Field Theory in
the $\alpha' \to 0$ limit and to its generalization to (A)dS
backgrounds, and conclude with a discussion of a simple local
compensator form of the field equations that displays the
unconstrained gauge symmetry of the non-local equations.
\\ \vskip 24pt
\begin{center}
{\it Based on the lectures presented by A. Sagnotti at the First
Solvay Workshop on Higher-Spin Gauge Theories held in Brussels on
May 12-14, 2004}
\end{center}
\vskip 5pt
\begin{flushleft}
\vspace{10pt} ULB-TH/04-22 \\ ROM2F-04/23
\end{flushleft}
\end{abstract}

\end{titlepage}
\newpage
\section{Introduction}

This article reviews some of the developments that led to the free
higher-spin equations introduced by Fang and Fronsdal
\cite{Fro78,Fro79} and the recent constructions of free non-local
geometric equations and local compensator forms of
\cite{Fra02a,Fra03,Sag04}. It is based on the lectures delivered
by A.~Sagnotti at the First Solvay Workshop, held in Brussels on
May 2004, carefully edited by the other authors for the online
Proceedings.

The theory of particles of arbitrary spin was initiated by Fierz
and Pauli in 1939 \cite{Fie39}, that followed a field theoretical
approach, requiring Lorentz invariance and positivity of the
energy. After the works of Wigner \cite{Wig39} on representations
of the Poincar\'{e} group, and of Bargmann and Wigner \cite{Bar48}
on relativistic field equations, it became clear that the
positivity of energy could be replaced by the requirement that the
one-particle states carry a unitary representation of the
Poincar\'{e} group. For massive fields of integer and half-integer
spin represented by totally symmetric tensors $\Phi_{\mu_1\dots
\mu_s}$ and $\Psi_{\mu_1\dots \mu_s}$, the former requirements are
encoded in the Fierz-Pauli conditions
\\
\begin{minipage}{0.5\textwidth}
\begin{eqnarray*}
(\square - M^2)\Phi_{\mu_1\dots \mu_s} &=& 0 \ , \nonumber \\
\d^{\mu_1} \Phi_{\mu_1\dots \mu_s} &=& 0 \ ,  \nonumber\\\nonumber
\end{eqnarray*}
\end{minipage}
\begin{minipage}{0.5\textwidth}
\begin{eqnarray}
(i\sd - M)\Psi_{\mu_1\dots \mu_s} &=& 0 \ , \label{masseq1} \\
\d^{\mu_1} \Psi_{\mu_1\dots \mu_s} &=& 0 \
.\label{masseq2}\\\nonumber
\end{eqnarray}
\end{minipage}
The massive field representations are also irreducible when a
($\gamma$-)trace condition
\begin{equation}
\eta^{\mu_1 \mu_2}\Phi_{\mu_1\mu_2\dots \mu_s} = 0 \ , \qquad
\gamma^{\mu_1}\Psi_{\mu_1\dots \mu_s} = 0 \ .
\end{equation}
is imposed on the fields.

A Lagrangian formulation for these massive spin $s$-fields was
first obtained in 1974 by Singh and Hagen \cite{Sin74},
introducing a sequence of auxiliary traceless fields of ranks
$s-2$, $s-3$, \dots 0 or 1, all forced to vanish when the field
equations are satisfied.

Studying the corresponding massless limit, in 1978 Fronsdal
obtained \cite{Fro78} four-dimensional covariant Lagrangians for
massless fields of any integer spin. In this limit, all the
auxiliary fields decouple and may be ignored, with the only
exception of the field of rank $s-2$, while the two remaining
traceless tensors of rank $s$ and $s-2$ can be combined into a
single tensor $\varphi_{\mu_1\cdots\mu_s}$ subject to the unusual
``double trace'' condition
\begin{equation}
\eta^{\mu_1\mu_2}\eta^{\mu_3\mu_4}\varphi_{\mu_1\cdots\mu_s}\ = \
0 \ .
\end{equation}
Fang and Fronsdal \cite{Fro79} then extended the result to
half-integer spins subject to the peculiar ``triple $\gamma$
trace'' condition
\begin{equation}
\gamma^{\mu_1}\gamma^{\mu_2} \gamma^{\mu_3}
\psi_{\mu_1\cdots\mu_s}\ = \ 0 \ .
\end{equation}

It should be noted that the description of massless fields in four
dimensions is particularly simple, since the massless irreducible
representations of the Lorentz group SO(3,1)$^\uparrow$ are
exhausted by totally symmetric tensors. On the other hand, it is
quite familiar from supergravity \cite{sugra} that in dimensions
$d > 4$ the totally symmetric tensor representations do not
exhaust all possibilities, and must be supplemented by mixed ones.
For the sake of simplicity, in this paper we shall confine our
attention to the totally symmetric case, focussing on the results
of \cite{Fra02a,Fra03,Sag04}. The extension to the mixed-symmetry
case was originally obtained in \cite{Bou02}, and will be reviewed
by C. Hull in his contribution to these Proceedings
\cite{chrisolvay}.

\section{From Fierz-Pauli to Fronsdal}

This section is devoted to some comments on the conceptual steps
that led to the Fronsdal \cite{Fro78} and Fang-Fronsdal
\cite{Fro79} formulations of the free high-spin equations. As a
first step, we describe the salient features of the Singh-Hagen
construction of the massive free field Lagrangians \cite{Sin74}.
For simplicity, we shall actually refer to spin 1 and 2 fields
that are to satisfy the Fierz-Pauli
conditions~(\ref{masseq1})-(\ref{masseq2}), whose equations are of
course known since Maxwell and Einstein. The spin-1 case is very
simple, but the spin-2 case already presents the key subtlety. The
massless limit will then illustrate the simplest instances of
Fronsdal gauge symmetries. The Kaluza-Klein mechanism will be also
briefly discussed, since it exhibits rather neatly the rationale
behind the Fang-Fronsdal auxiliary fields for the general case. We
then turn to the novel features encountered with spin 3 fields,
before describing the general Fronsdal equations for massless
spin-$s$ bosonic fields. The Section ends with the extension to
half-integer spins.

\subsection{Fierz-Pauli conditions}

Let us first introduce a convenient compact notation. Given a
totally symmetric tensor $\varphi$, we shall denote by $\partial
\varphi$, $\partial \cdot \varphi$ and $\varphi^\prime$ (or, more
generally, $\varphi^{[p]}$) its gradient, its divergence and its
trace (or its $p$-th trace), with the understanding that in all
cases the implicit indices are totally symmetrized.

Singh and Hagen \cite{Sin74} constructed explicitly Lagrangians
for spin-$s$ fields that give the correct Fierz-Pauli conditions.
For spin 1 fields, their prescription reduces to the Lagrangian
\begin{equation}
{\cal L}_{spin 1} = - \frac{1}{2} (\partial_\mu \Phi_{\nu} )^2 -
\frac{1}{2} (\partial \cdot \Phi)^2 - \frac{M^2}{2} ( \Phi_\mu)^2
\ ,
\end{equation}
that gives the Proca equation
\begin{equation}\Box\Phi_{\mu}-\partial_{\mu}(\partial\cdot\Phi)-M^{2}\Phi_{\mu}=0
\ .\label{proca}
\end{equation}
Taking the divergence of this field equation, one obtains
immediately $\d^{\mu} \Phi_{\mu}=0$, the Fierz-Pauli
transversality condition (\ref{masseq2}), and hence the
Klein-Gordon equation for $\Phi_\mu$.

In order to generalize this result to spin-2 fields, one can begin
from
\begin{equation}
\label{spin2} {\cal L}_{spin 2} = - \frac{1}{2} (\partial_\mu
\Phi_{\nu\rho} )^2 + \frac{\alpha}{2} (\partial \cdot \Phi_\nu)^2
- \frac{M^2}{2} ( \Phi_{\mu\nu})^2 \ ,
\end{equation}
where the field $\Phi_{\mu\nu}$ is traceless. The corresponding
equation of motion reads
\begin{equation}
\Box\Phi_{\mu\nu}-\frac{\alpha}{2} \left( \partial_{\mu} \partial\cdot\Phi_\nu +
\partial_{\nu}\partial\cdot\Phi_\mu - \frac{2}{D} \; \eta_{\mu\nu} \;\partial \cdot \partial \cdot
\Phi \right)-M^{2}\Phi_{\mu\nu}=0 \ , \label{eqSpin2}
\end{equation}
whose divergence implies
\begin{equation}
\left(1 - \frac{\alpha}{2} \right) \Box \partial \cdot \Phi_\nu +
\alpha \left(- \frac{1}{2} + \frac{1}{D} \right) \partial_\nu
\partial \cdot \partial \cdot \Phi -  M^2 \partial \cdot \Phi_\nu
= 0 \ .
\end{equation}
Notice that, in deriving these equations, we have made an essential use of
the condition that $\Phi$ be traceless.

In sharp contrast with the spin 1 case, however, notice that now
the transversality condition is not recovered. Choosing $\alpha 2$ would eliminate some terms, but one would still need the
additional constraint $\d\cdot \d\cdot \Phi = 0$. Since this is
not a consequence of the field equations, the naive system
described by $\Phi$ and equipped with the Lagrangian $\cL_{spin2}$
is unable to describe the free spin 2 field.

One can cure the problem introducing an auxiliary scalar field
$\pi$ in such a way that the condition $\d\cdot \d\cdot \Phi =0$
be a consequence of the Lagrangian. Let us see how this is the
case, and add to (\ref{spin2}) the term
\begin{equation}
{\cal L}_{add} = \pi\; \partial \cdot \partial \cdot \Phi + c_1
(\partial_\mu \pi)^2 + c_2 \pi^2\ ,
\end{equation}
where $c_{1,2}$ are a pair of constants. Taking twice the
divergence of the resulting equation for $\Phi_{\mu\nu}$ gives
\begin{equation}
\left[ \left(2 - D \right)\square - D M^2 \right] \partial \cdot \partial
\cdot \Phi + \left( D - 1 \right) \Box^2 \pi = 0 \ ,\label{syst1}
\end{equation}
while the equation for the auxiliary scalar field reduces to
\begin{equation}
\partial \cdot \partial \cdot \Phi + 2( c_2 - c_1 \Box) \pi \ = \ 0 \ .  \label{syst2}
\end{equation}
Eqs. (\ref{syst1}) and (\ref{syst2}) can be regarded as a linear
homogeneous system in the variables $\d\cdot \d\cdot \Phi$ and
$\pi$. If the associated determinant never vanishes, the only
solution will be precisely the missing condition $\d\cdot \d\cdot
\Phi = 0$, together with the condition that the auxiliary field
vanish as well, $\pi = 0$, and as a result the transversality
condition will be recovered. The coefficients $c_1$ and $c_2$ are
thus determined by the condition that the determinant of the
system
\begin{eqnarray}
\Delta &=& -2DM^2c_2 + 2((2-D)c_2+DM^2c_1)\square \\
&-& (2(2-D)c_1-(D-1))\square\square
\end{eqnarray}
be algebraic, {\it i.e.} proportional to the mass $M$ but without
any occurrence of the D'Alembert operator $\square$. Hence, for $D
> 2$,
\begin{equation}
 c_1 = \frac{(D-1)}{2(D-2)} \ , \qquad \quad
c_2 = \frac{M^2 D (D-1)}{2(D-2)^2} \ .
\end{equation}
The end conclusion is that the complete equations imply
\begin{eqnarray}
& \pi = 0 \ , \qquad &\partial \cdot \partial \cdot \Phi = 0\ ,  \\
& \partial \cdot \Phi_\nu = 0  \ , \qquad &\Box \Phi_{\mu\nu} -
M^2 \Phi_{\mu\nu} = 0 \ ,
\end{eqnarray}
the Fierz-Pauli conditions (\ref{masseq1}) and (\ref{masseq2}), so
that the inclusion of a single auxiliary scalar field leads to an
off-shell formulation of the free massive spin-2 field.

\subsection{``Fronsdal'' equation for spin 2}

We can now take the $M \rightarrow 0$ limit, following in spirit
the original work of Fronsdal \cite{Fro78}. The total Lagrangian
$\cL_{spin 2} + \cL_{add} $ then becomes
\begin{equation}
{\cal L} \ =\  - \, \frac{1}{2} \,(\partial_\mu \Phi_{\nu\rho} )^2
+ (\partial \cdot \Phi_\nu)^2 + \pi \partial \cdot \partial \cdot
\Phi + \frac{D-1}{2(D-2)}\, (\partial_\mu \pi)^2 \ ,
 \label{Lmassless}
\end{equation}
whose equations of motion are
\begin{eqnarray}
&& \Box\Phi_{\mu\nu}-  \partial_{\mu} \partial\cdot\Phi_\nu -
\partial_{\nu}\partial\cdot\Phi_\mu + \frac{2}{D} \; \eta_{\mu\nu}
\;\partial \cdot \partial \cdot
\Phi  + \partial_\mu \partial_\nu \pi=0 \ , \label{eq1massless} \\
&& \frac{D-1}{D-2} \, \Box \pi - \partial \cdot \partial \cdot
\Phi = 0 \ . \label{eq2massless}
\end{eqnarray}

The representation of the massless spin 2-gauge field via a
traceless two-tensor $\Phi_{\mu\nu}$ and a scalar $\pi$ may seem a
bit unusual. In fact, they are just an unfamiliar basis of fields,
and the linearized Einstein gravity in its standard form is simply
recovered once they are combined in the unconstrained two-tensor
\begin{equation}\varphi_{\mu\nu} = \Phi_{\mu\nu} + \frac{1}{D-2} \; \eta_{\mu\nu} \;
\pi \ .
\end{equation}
In terms of $\varphi_{\mu\nu}$, the field equations and the
corresponding gauge transformations then become
\begin{eqnarray}
&& {\cal F}_{\mu\nu} \equiv \Box \varphi_{\mu\nu} - (\partial_\mu \partial
\cdot \varphi_\nu + \partial_\nu \partial \cdot \varphi_\mu) + \partial_\mu
\partial_\nu \varphi' = 0 \ , \\
&& \delta \varphi_{\mu\nu} = \partial_\mu \Lambda_\nu +
\partial_\nu \Lambda_\mu \ ,
\end{eqnarray}
that are precisely the linearized Einstein equations, where the
``Fronsdal operator'' $\cF_{\mu\nu}$ is just the familiar Ricci
tensor. The corresponding Lagrangian reads
\begin{equation}{\cal L} = - \frac{1}{2} \left( \partial_\mu \varphi_{\nu\rho} \right)^2 +
\left( \partial \cdot \varphi_\mu \right)^2 + \frac{1}{2} \left(
\partial_\mu \varphi' \right)^2 + \varphi' \partial \cdot \partial \cdot
\varphi \ ,
\end{equation}
and yields the Einstein equations $\cF_{\mu\nu} - \frac{1}{2}
\eta_{\mu\nu}\cF^\prime = 0$, that only when combined with their
trace imply the previous equation, $\cF_{\mu\nu} = 0$.

The massless case is very interesting by itself, since it exhibits
a relatively simple instance of gauge symmetry, but also for
deducing the corresponding massive field equations via a proper
Kaluza-Klein reduction. This construction, first discussed in
\cite{massiveKK}, is actually far simpler than the original one of
\cite{Sin74} and gives a rationale to their choice of auxiliary
fields.

Let us content ourselves with illustrating the Kaluza-Klein
mechanism for spin 1 fields. To this end, let us introduce a
field\footnote{Capital Latin letters denote here indices in $D+1$
dimensions, while Greek letters denote the conventional ones in
$D$ dimensions.} $A_M$ living in $D+1$ dimensions, that decomposes
as $A_M = (A_\mu(x,y), \pi(x,y))$, where $y$ denotes the
coordinate along the extra dimension. One can expand these
functions in Fourier modes in $y$ and a single massive mode
corresponding to the $D$-dimensional mass $m$, letting for
instance $A_M = (A_\mu(x), -i\pi(x))\, \exp{i m y}$ where the
judicious insertion of the factor $-i$ will ensure that the field
$\pi$ be real. The $D+1$-dimensional equation of motion and gauge
transformation
\begin{eqnarray}
&& \Box A_M - \partial_M \partial \cdot A = 0 \ , \\
&& \delta A_M = \partial_M \Lambda
\end{eqnarray}
then determine the $D$-dimensional equations
\begin{eqnarray}
&& \left( \Box - m^2 \right) A_\mu - \partial_\mu \left( \partial \cdot A + m \pi \right)=0 ,\nonumber \\
&& \left( \Box - m^2 \right) \pi + m \left( \partial \cdot A + m \pi  \right) = 0 \ ,  \\
&& \delta A_\mu = \partial_\mu \Lambda,  \qquad \delta \pi = - m
\Lambda \ , \nonumber
\end{eqnarray}
where the leftover massive gauge symmetry, known as a Stueckelberg
symmetry, is inherited from the higher dimensional gauge symmetry.
Fixing the gauge so that $\pi = 0$, one can finally recover the
Proca equation~(\ref{proca}) for $A_\mu$. The spin-2 case is
similar, and the proper choice is $\varphi_{MN}
(\varphi_{\mu\nu},-i\varphi_\mu,-\varphi)\,\exp{imy}$, so that the
resulting gauge transformations read
\begin{eqnarray}
&& \delta \varphi_{MN} = \partial_M \Lambda_N + \partial_N \Lambda_M \nonumber \ , \\
&& \delta \varphi_{\mu\nu} = \partial_\mu \Lambda_\nu + \partial_\nu \Lambda_\mu \nonumber \ , \\
&&  \delta \varphi_{\mu} = \partial_\mu \Lambda - m \Lambda_\mu  \ ,\\
&& \delta \varphi = - 2 m \Lambda \ .\nonumber
\end{eqnarray}
In conclusion, everything works as expected when the spin is lower
than or equal to two, and the Fierz-Pauli conditions can be easily
recovered. However, some novelties do indeed arise when then spin
becomes higher than two.

Let us try to generalize the theory to spin-3 fields by insisting
on the equations
\begin{eqnarray}
&& {\cal F}_{\mu\nu\rho} \equiv \Box \varphi_{\mu\nu\rho} - (\partial_\mu
\partial \cdot \varphi_{\nu\rho} + {perm}\,)
 + (\partial_\mu \partial_\nu \varphi'_\rho +{perm}) = 0 \ , \label{eq3} \\
&& \delta \varphi_{\mu\nu\rho} = \partial_\mu \Lambda_{\nu\rho} +
\partial_\nu \Lambda_{\rho\mu} + \partial_\rho \Lambda_{\mu\nu}\ , \label{gauge3}
\end{eqnarray}
that follow the same pattern, where \textit{perm} denotes cyclic
permutations of $\mu\nu\rho$. In this formulation, there are no
auxiliary fields and the trace $\varphi^\prime$ of the gauge field
does not vanish.

Let us first remark that, under a gauge transformation, $\cF$
transforms according to
\begin{equation}
\delta \cF_{\mu_1\mu_2\mu_3} = 3 \d_{\mu_1}\d_{\mu_2}\d_{\mu_3}
\Lambda^\prime \ . \label{novanish}
\end{equation}
Therefore, \emph{$\cF$ is gauge invariant if and only if the gauge
parameter is traceless, $\Lambda^\prime = 0$}. This rather
unnatural condition will recur systematically for all higher
spins, and will constitute a drawback of the Fronsdal formulation.

We can also see rather neatly the obstruction to a geometric gauge
symmetry of the spin-3 Fronsdal Lagrangian along the way inspired
by General Relativity. Indeed, the spin-3 Fronsdal equation
differs in a simple but profound way from the two previous cases,
since both for spin 1 and for spin 2 \emph{all lower-spin
constructs} built out of the gauge fields are present, while for
spin 3 only constructs of spin 3 ($\varphi_{\mu\nu\rho}$), spin 2
($\partial \cdot \varphi_{\mu\nu}$) and spin 1 ($\varphi'_\mu$)
are present. Actually, de Wit and Freedman \cite{deW80} classified
long ago the higher-spin analogs of the spin-2 Christoffel
connection $\Gamma_{mu;\nu_1\nu_2}$: they are a hierarchy of
connections $\Gamma_{\mu_1 \ldots \mu_k;\nu_1 \nu_2 \nu_3}$, with
$k=1, \ldots (s-1)$, that contain $k$ derivatives of the gauge
field. They also noticed that the analog of the Riemann tensor,
that for spin 3 would be $\Gamma_{\mu_1 \mu_2 \mu_3;\nu_1 \nu_2
\nu_3}$, would in general contain $s$ derivatives of the gauge
field, and \emph{related the Fronsdal operator $\cF$ to the trace
of the second connection}.

One way to bypass the problem is to construct a field equation
with two derivatives depending on the true Einstein tensor for
higher spins, that however, as we have anticipated, contains $s$
derivatives in the general case. This can be achieved, but
requires that non-local terms be included both in the field
equations and in the Lagrangian. This approach will be further
developed in section~\ref{nonlocal}. Another option is to
compensate the non-vanishing term in the right-hand side
of~(\ref{novanish}) by introducing a new field, a
\emph{compensator}. As we shall see, this possibility is actually
suggested by String Theory, and will be explained in
section~\ref{tripletscomp}.

But before describing these new methods, let us first describe the
general Fronsdal formulation for arbitrary spin. In this way we
shall clearly identify the key role of the \emph{trace condition
on the gauge parameter} that we already encountered for spin 3 and
of the \emph{double trace condition on the field}, that will first
present itself for spin 4.

\subsection{Fronsdal equations for arbitrary integer spin}

We can now generalize the reasoning of the previous paragraph to
arbitrary integer spins. Since we shall use extensively the
compact notation of \cite{Fra02a}, omitting all indices, is it
useful to recall the following rules:
\begin{eqnarray}
&& \left( \partial^{\; p} \ \varphi  \right)^{\; \prime} \ = \ \Box \
\partial^{\; p-2} \
\varphi \ + \ 2 \, \partial^{\; p-1} \  \partial \cdot \varphi \ + \
\partial^{\; p} \
\varphi {\;'} \nonumber \ , \\
&& \partial^{\; p} \ \partial^{\; q} \ = \ \left( {p+q} \atop p \right) \
\partial^{\; p+q} \nonumber \ ,
\\
&& \partial \cdot  \left( \partial^{\; p} \ \varphi \right) \ = \ \Box \
\partial^{\; p-1} \ \varphi \ + \
\partial^{\; p} \ \partial \cdot \varphi \label{notation} \ , \\
&& \partial \cdot  \eta^{\;k} \ = \ \partial \, \eta^{\;k-1} \nonumber ,\\
&& \left( \eta^k \, T_{(s)} \,  \right)^\prime \ = \ k \, \left[
\, {\cal D} \, + \, 2(s+k-1) \,  \right]\, \eta^{k-1} \, T_{(s)} \
+ \ \eta^k \, T_{(s)}^\prime \ .\nonumber
\end{eqnarray}
In this compact form, the generic Fronsdal equation and its gauge
transformations read simply
\begin{eqnarray}
&& {\cal F} \equiv \Box \varphi - \partial \partial \cdot \varphi +
\partial^2 \varphi'=0 \ ,
 \label{fronseq} \\
&& \delta \varphi = \partial \Lambda \ .
\end{eqnarray}
In order to find the effect of the gauge transformations on the
Fronsdal operators $\cF$, one must compute the terms
\begin{eqnarray*}
&&\quad \delta (\partial \cdot \varphi ) = \Box \Lambda + \partial
\partial \cdot \Lambda \ ,
\\&&\quad \delta \varphi' = 2 \partial \cdot \Lambda + \partial \Lambda' \nonumber \ ,
\end{eqnarray*}
and the result is, for arbitrary spin,
\begin{eqnarray}
\delta {\cal F} = \Box ( \partial \Lambda) - \partial ( \Box \Lambda +
\partial \partial \cdot \Lambda ) + \partial^2 ( 2 \partial \cdot \Lambda
+ \partial \Lambda' ) = {  3 \partial^3 \Lambda'} \ ,
\end{eqnarray}
where we used $-\partial (\partial \partial\cdot \Lambda)= -2\partial^2
(\partial\cdot\Lambda)$, and $\partial^2\partial\Lambda^\prime = 3
\partial^3 \Lambda^\prime$. Therefore, in all cases
where $\Lambda'$ does not vanish identically, {\it i.e.} for spin $s \geq
3$, the gauge invariance of the equations requires that the gauge
parameter be traceless
\begin{equation}
\Lambda^\prime = 0 \ .\label{lambda0}
\end{equation}
As second step, one can derive the Bianchi identities for all
spins computing the terms
\begin{eqnarray}
&& \partial \cdot {\cal F} = \Box \partial \varphi' - \partial \partial
\cdot \partial \cdot \varphi + \partial^2 \partial \cdot \varphi' \nonumber, \\
&& {\cal F}' = 2\Box \varphi' - 2 \partial \cdot \partial \cdot \varphi +
\partial^2 \varphi'' + \partial \partial \cdot \varphi' \nonumber, \\
&& \partial {\cal F}' = 2\Box \partial \varphi' - 2 \partial \; \partial
\cdot \partial \cdot \varphi + 3 \partial^3 \varphi'' + 2 \partial^2
\partial \cdot \varphi' \nonumber.
\end{eqnarray}
Therefore, the Fronsdal operator $\cF$ satisfies in general the
"anomalous" Bianchi identities
\begin{equation}
\partial \cdot{\cal F} - \frac{1}{2} \partial {\cal F}'-\frac{3}{2}\partial^3\varphi^{\prime\prime} \ ,\label{bianchi}
\end{equation}
where the additional term on the right first shows up for spin
$s=4$. In the Fronsdal construction, one is thus led to impose the
constraint $\varphi^{\prime\prime} = 0$ for spins $s\geq 4$, since
the Lagrangians would vary according to
\begin{equation}
\delta {\cal L} \ = \ \delta \varphi \left( \cF - \frac{1}{2} \eta
\cF' \right) \ ,
\end{equation}
that does not vanish unless the double trace of $\varphi$ vanishes
identically. Indeed,
\begin{equation}
\partial \cdot \left( \cF - \frac{1}{2} \cF' \right) = -\frac{3}{2} \partial^3 \varphi^{\prime\prime}-
\half\eta \partial \cdot \cF^\prime \ , \label{eqBianc}
\end{equation}
where the last term gives a vanishing contribution to $\delta
{\cal L}$ if the parameter $\Lambda$ is traceless. To reiterate,
this relation is at the heart of the usual restrictions present in
the Fronsdal formulation to traceless gauge parameters and doubly
traceless fields, needed to ensure that the variation of the
lagrangian
\begin{equation}
\delta \cL = \delta \varphi\, \cG \label{deltaL},
\end{equation}
where $\cG = \cF - \half \eta \cF^\prime$, vanishes for
$\delta\varphi = \partial \Lambda$.

We can also extend the Kaluza-Klein construction to the spin-$s$ case.
Given the double trace condition $\varphi^{\prime\prime}=0$, the reduction
$\varphi^{(s)}_{D+1}$ from $D+1$ dimensions to $D$ dimensions gives rise
to the tensors $\varphi_D^{(s)}$, \dots, $\varphi_D^{(s-3)}$ of rank $s$
to $s-3$ only. In addition, the trace condition on the gauge parameter
implies that only two tensors $\Lambda_D^{(s-1)}$ and $\Lambda_D^{(s-2)}$
are generated in $D$ dimensions. Gauge fixing the Stueckelberg symmetries,
one is left with only two traceful fields $\varphi_D^{(s)}$ and
$\varphi_D^{(s-3)}$. But a traceful spin-$s$ tensor contains traceless
tensors of ranks $s$, $s-2$, $s-4$, etc. Hence, the two remaining fields
$\varphi_D^{(s)}$ and $\varphi_D^{(s-3)}$ contain precisely the tensors
introduced by Singh and Hagen \cite{Sin74}: a single traceless tensor for
all ranks from $s$ down to zero, with the only exception of the
rank-$(s-1)$ tensor, that is missing.

\subsection{Massless fields of half-integer spin}
\label{chapFerm} Let us now turn to the fermionic fields, and for
simplicity let us begin with the Rarita-Schwinger equation
\cite{RS}, familiar from supergravity \cite{supergravity}
\begin{equation}\gamma^{\mu\nu\rho} \, \partial_\nu \psi_\rho = 0 \ ,
\end{equation}
that is invariant under the gauge transformation
\begin{equation} \delta \psi_\mu = \d_\mu \eps \ , \end{equation}
where $\gamma^{\mu\nu\rho}$ denotes the fully antisymmetric
product of three $\gamma$ matrices, normalized to their product
when they are all different:
\begin{equation}
\gamma^{\mu\nu\rho} \ = \ \gamma^\mu \gamma^{\nu\rho}
 - \eta^{\mu\nu} \gamma^\rho +  \eta^{\mu\rho} \gamma^\nu
 \ . \end{equation}
Contracting the Rarita-Schwinger equation with $\gamma_\mu$ yields
\begin{equation}
\gamma^{\nu\rho}\d_\nu \psi_\rho = 0 \ ,
\end{equation}
and therefore the field equation for spin 3/2 can be written in
the alternative form
\begin{equation}
\sd \psi_\mu - \d_\mu \psis = 0 \ .
\end{equation}

Let us try to obtain a similar equation for a spin-$5/2$ field,
defining the Fang-Fronsdal operator $S_{\mu\nu}$, in analogy with
the spin-$3/2$ case, as
\begin{eqnarray}
S_{\mu\nu} \equiv i\; \left( \sd \psi_{\mu\nu} - \d_\mu \psis_\nu
- \d_\nu \psis_\mu \right) \ = \ 0 \ ,
\end{eqnarray}
and generalizing naively the gauge transformation to
\begin{equation}
\delta \psi_{\mu\nu} =  \d_\mu \eps_{\nu} + \d_\nu \eps_\mu \ .
\end{equation}
Difficulties similar to those met in the bosonic case for spin 3
readily emerge: this equation is not gauge invariant, but
transforms as
\begin{equation} \delta S_{\mu\nu} = -2 i\; \d_\nu\d_\mu \epss \
,\end{equation} and a similar problem will soon arise with the
Bianchi identity. However, as in the bosonic case, following Fang
and Fronsdal \cite{Fro79}, one can constrain both the fermionic
field and the gauge parameter $\eps$ so that the gauge symmetry
hold and the Bianchi identity take a non-anomalous form.

We can now consider the generic case of half-integer spin $s +
1/2$ \cite{Fro79}. In the compact notation (\ref{notation}) the
equation of motion and the gauge transformation read simply
\begin{eqnarray}
&&S \ \equiv \ i \; (\sd \psi - \d \psis) \ = \ 0 \ ,\\
&&\delta \psi = \d \eps \ .
\end{eqnarray}
Since
\begin{equation} \delta S = -2 i \d^2 \epss \ ,
\end{equation}
to ensure the gauge invariance of the field equation, one must
demand that the gauge parameter be {\it $\gamma$-traceless},
\begin{equation}
\epss = 0 \ . \label{tracegauge}
\end{equation}

Let us now turn to the Bianchi identity, computing
\begin{equation}
\d \cdot S - \half \d S^\prime - \half \sd\, \Slash S \ ,
\end{equation}
where the last term contains the $\gamma$-trace of the operator.
It is instructive to consider in detail the individual terms. The
trace of $S$ is
\begin{equation}
S^\prime = i (\sd \psi^\prime - 2 \d \cdot \Slash \psi - \d \Slash
{\psi}^\prime ) \ ,
\end{equation}
and therefore, using the rules in (\ref{notation})
\begin{equation}
- \half \d S^\prime = - \frac{i}{2}(\sd \d \psi^\prime - 2 \d \d
\cdot \Slash \psi - 2 \d^2 \Slash {\psi}^\prime) \ .
\end{equation}
Moreover, the divergence of $S$ is
\begin{equation}
\d \cdot S = i (\sd \d \cdot \psi - \square \Slash \psi - \d \d
\cdot \Slash \psi) \ .
\end{equation}
Finally, from the $\gamma$ trace of $S$
\begin{equation}
\Slash S = i (-2 \sd \Slash \psi + 2 \d \cdot \psi - \d
\psi^\prime) \ ,
\end{equation}
one can obtain
\begin{equation}
- \half \sd \Slash S = - \frac{i}{2} (- 2 \square \Slash \psi + 2
\sd \d \cdot \psi - \d \sd \psi^\prime) \ ,
\end{equation}
and putting all these terms together yields the Bianchi identity
\begin{equation}
\d \cdot S - \half \d S^\prime - \half \sd \Slash S = i \d^2 \Slash
{\psi}^\prime \ .\label{bianchihalf}
\end{equation}
As in the bosonic case, this identity contains an ``anomalous''
term that first manifests itself for spin $s=7/2$, and therefore
one is lead in general to impose the ``triple $\gamma$-trace''
condition
\begin{equation}
\Slash{\psi}^\prime = 0 \ .\label{tracepsi}
\end{equation}

One can also extend the Kaluza-Klein construction to the spin $s +
1/2$ in order to recover the massive Singh-Hagen formulation. The
reduction from $D+1$ dimensions to $D$ dimensions will turn the
massless field $\psi^{(s)}_{D+1}$ into massive fields of the type
$\psi^{(s)}_{D}, \psi^{(s-1)}_{D}$ and $\psi^{(s-2)}_{D}$, while
no lower-rank fields can appear because of the triple
$\gamma$-trace condition~(\ref{tracepsi}). In a similar fashion,
the gauge parameter $\eps_{D+1}^{(s-1)}$ reduces only to a single
field $\eps_{D}^{(s-1)}$, as a result of the $\gamma$-trace
condition~(\ref{tracegauge}). Gauge fixing the Stueckelberg
symmetries one is finally left with only two fields
$\psi^{(s)}_{D}$ and $\psi^{(s-2)}_{D}$ that contain precisely the
$\gamma$-traceless tensors introduced by Singh and Hagen in
\cite{Sin74}.

\section{Free non-local geometric equations}
\label{nonlocal} In the previous section we have seen that is
possible to construct a Lagrangian for higher-spin bosons imposing
the unusual Fronsdal constraints
\begin{equation}
\Lambda^\prime = 0, \qquad \varphi^{\prime\prime} = 0
\end{equation}
on the fields and on the gauge parameters. Following
\cite{Fra02a,Fra03}, we can now construct higher-spin gauge
theories with unconstrained gauge fields and parameters.

\subsection{Non-local Fronsdal-like operators}

We can motivate the procedure discussing first in some detail the
relatively simple example of a spin-3 field, where
\begin{equation}
\delta \cF_{\mu\nu\rho} = 3 \; \d_\mu \d_\nu \d_\rho
\Lambda^\prime \ .
\end{equation}
Our purpose is to build a non-local operator $\cF_{NL}$ that
transforms exactly like the Fronsdal operator $\cF$, since the
operator $\cF - \cF_{NL}$ will then be gauge invariant without any
additional constraint on the gauge parameter. One can find rather
simply the non-local constructs
\begin{eqnarray}
&&\frac{1}{3\square} \, [\d_\mu \d_\nu \cF^\prime_\rho + \d_\nu
\d_\rho \cF^\prime_\mu
+\d_\rho \d_\mu \cF^\prime_\nu ] \ ,\\
&&\frac{1}{3\square} \, [\d_\mu \d \cdot \cF^\prime_{\nu\rho} +
\d_\nu \d \cdot
\cF^\prime_{\rho\mu} + \d_\rho \d \cdot \cF^\prime_{\mu\nu} ]\ ,\\
&&\frac{1}{\square^2} \, \d_\mu\d_\nu\d_\rho \d \cdot \cF^\prime \
,
\end{eqnarray}
but the first two expressions actually coincide, as can be seen
from the Bianchi identities~(\ref{bianchi}), and as a result one
is led to two apparently distinct non-local fully gauge invariant
field equations
\begin{eqnarray}
&& \cF_{\mu_1\mu_2\mu_3} - \frac{1}{3\square} \ [\d_{\mu_1}
\d_{\mu_2} \cF^\prime_{\mu_3} + \d_{\mu_2} \d_{\mu_3}
\cF^\prime_{\mu_1} +\d_{\mu_3} \d_{\mu_1} \cF^\prime_{\mu_2} ] = 0 \ , \label{gener} \\
&& \cF^{new}_{\mu_1\mu_2\mu_3} \equiv \cF_{\mu_1\mu_2\mu_3} -
\frac{1}{\square^2}\, \d_{\mu_1}\d_{\mu_2}\d_{\mu_3}\d \cdot
\cF^\prime = 0 \ .
\end{eqnarray}
These equations can be actually turned into one another, once they
are combined with their traces, but the second form, which we
denote by $\cF^{new}$, is clearly somewhat simpler, since it rests
on the addition of the single scalar construct $\d \cdot
\cF^\prime$. From $\cF^{new}$, one can build in the standard way
\begin{equation}
G_{\mu_1\mu_2\mu_3} \equiv \cF^{new}_{\mu_1\mu_2\mu_3} - \half
(\eta_{\mu_1\mu_2} {\cF^{new}_{\mu_3}}^\prime + \eta_{\mu_2\mu_3}
{\cF^{new}_{\mu_1}}^\prime +\eta_{\mu_3\mu_1}
{\cF^{new}_{\mu_2}}^\prime) \ ,
\end{equation}
and one can easily verify that
\begin{equation}
\d \cdot G_{\mu_1\mu_2} = 0 \ .
\end{equation}
This identity suffices to ensure the gauge invariance of the
Lagrangian. Moreover, we shall see that in all higher-spin cases,
the non-local construction will lead to a similar, if more
complicated, identity, underlying a Lagrangian formulation that
does not need any double trace condition on the gauge field. It is
worth stressing this point: we shall see that, modifying the
Fronsdal operator in order to achieve gauge invariance without any
trace condition on the parameter, the Bianchi identities will
change accordingly and the ``anomalous'' terms will disappear,
leading to corresponding gauge invariant Lagrangians.

Returning to the spin-3 case, one can verify that the Einstein
tensor $G_{\mu_1\mu_2\mu_3}$ follows from the Lagrangian
\begin{eqnarray}
\cL &=& -\half (\d_\mu \varphi_{\mu_1\mu_2\mu_3})^2 + \frac{3}{2}
(\d \cdot \varphi_{\mu_1\mu_2})^2 - \frac{3}{2} (\d \cdot
\varphi^\prime)^2 + \frac{3}{2} (\d_\mu \varphi_{\mu_1}^\prime)^2
\nonumber \\ &&+ 3 \varphi^\prime_{\mu_1} \d\cdot\d\cdot
\varphi_{\mu_1} + 3 \div\div\div \varphi \frac{1}{\square}\div
\varphi^\prime - \div\div\div\varphi \frac{1}{\square^2} \div
\div\div \varphi \ ,
\end{eqnarray}
that is fully gauge invariant under
\begin{equation} \delta \varphi_{\mu_1\mu_2\mu_3} = \d_{\mu_1} \Lambda_{\mu_2\mu_3} +
\d_{\mu_2} \Lambda_{\mu_3\mu_1} + \d_{\mu_3} \Lambda_{\mu_1\mu_2}
\ .
\end{equation}

For all higher spins, one can arrive at the proper analogue of
(\ref{gener}) via a sequence of pseudo-differential operators,
defined recursively as
\begin{equation}
\cF^{(n+1)} = \cF^{(n)} +
\frac{1}{(n+1)(2n+1)}\frac{\d^2}{\square} {\cF^{(n)}}^\prime -
\frac{1}{n+1} \frac{\d}{\square}\,\div \cF^{(n)} \ ,
\end{equation}
where the initial operator $\cF^{(1)} = \cF$ is the classical
Fronsdal operator. The gauge transformations of the $\cF^{(n)}$,
\begin{equation}
\delta \cF^{(n)} = (2n+1)\frac{\d^{2n+1}}{\square^{n-1}}
\Lambda^{[n]} \ ,
\end{equation}
involve by construction higher traces of the gauge parameter.
Since the $n$-th trace $\Lambda^{[n]}$ vanishes for all $n
> (s - 1)/2$, the first corresponding operator $\cF^{(n)}$ will
be gauge invariant without any constraint on the gauge parameter.
A similar inductive argument determines the Bianchi identities for
the $\cF^{(n)}$,
\begin{equation}
\div\cF^{(n)} - \frac{1}{2n} \d {\cF^{(n)}}^\prime = -\left( 1+
\frac{1}{2n}\right) \frac{\d^{2n+1}}{\square^{n-1}}\varphi^{[n+1]}
\ ,
\end{equation}
where the \emph{anomalous} contribution depends on the $(n+1)$-th
trace $\varphi^{[n+1]}$ of the gauge field, and thus vanishes for
$n > (s/2 - 1)$.

The Einstein-like tensor corresponding to $\cF^{(n)}$
\begin{equation}
G^{(n)} = \sum_{p=0}^{n-1} \frac{(-1)^p(n-p)!}{2^p
n!}\eta^p\cF^{(n)[p]} \ ,
\end{equation}
is slightly more complicated than its lower-spin analogs, since it
involves in general multiple traces, but an inductive argument
shows that
\begin{equation}
\partial \cdot G^{(n)} = 0 \ ,
\end{equation}
so that $G^{(n)}$ follows indeed from a Lagrangian of the type
\begin{equation}
{\cal L} \sim \varphi \, G^{(n)} \ .
\end{equation}
We shall soon see that $G^{(n)}$ has a very neat geometrical
meaning. Hence, the field equations, not directly the Lagrangians,
are fully geometrical in this formulation.

\subsection{Geometric equations}

Inspired by General Relativity, we can reformulate the non-local
objects like the Ricci tensor $\cF^{(n)}$ introduced in the
previous section in geometrical terms. We have already seen that,
following de Wit and Freedman \cite{deW80}, one can define
generalized connections and Riemann tensors of various orders in
the derivatives for all spin-$s$ gauge fields as extensions of the
spin-2 objects as
\begin{eqnarray*}
\Gamma_{\mu ; \nu_1 \nu_2}\quad \Rightarrow \quad \Gamma_{\mu_1\dots
\mu_{s-1};\nu_1\dots\nu_s}\ , \\
R_{\mu_1\mu_2 ; \nu_1 \nu_2}\quad \Rightarrow \quad R_{\mu_1\dots
\mu_{s};\nu_1\dots\nu_s} \ ,
\end{eqnarray*}
and actually a whole hierarchy of connections $\Gamma_{\mu_1\dots
\mu_{k};\nu_1\dots\nu_s}$ whose last two members are the
connection and the curvature above. In order to appreciate better
the meaning of this generalization, it is convenient to recall
some basic facts about linearized Einstein gravity. If the metric
is split according to $g_{\mu\nu} = \eta_{\mu\nu} + h_{\mu\nu}$,
the condition that $g$ be covariantly constant leads to the
following relation between its deviation $h$ with respect to flat
space and the linearized Christoffel symbols:
\begin{equation}
\d_\rho h_{\mu\nu} = \Gamma_{\nu;\rho\mu} + \Gamma_{\mu;\rho\nu} \
.
\end{equation}
In strict analogy, the corresponding relation for spin 3 is
\begin{equation}
\d_\sigma \d_\tau \varphi_{\mu\nu\rho} \ = \
\Gamma_{\nu\rho;\sigma\tau\mu}+ \Gamma_{\rho\mu;\sigma\tau\nu} +
\Gamma_{\mu\nu;\sigma\tau\rho} \ .
\end{equation}

It is possible to give a compact expression for the connections of
\cite{deW80} for arbitrary spin,
\begin{equation}
\Gamma^{(s-1)} = \frac{1}{s} \sum_{k=0}^{s-1} \frac{(-1)^k}{\left(
s-1\atop k \right) } \ \d^{s-k-1} \nabla^k \varphi \ ,
\end{equation}
where the derivatives $\nabla$ carry indices originating from the
gauge field. This tensor is actually the proper analogue of the
Christoffel connection for a spin-$s$ gauge field, and transforms
as
\begin{equation}
\delta \Gamma_{\alpha_1 \cdots \alpha_{s-1};\beta_1 \cdots
\beta_s} \ \sim \ \partial_{\beta_1} \cdots \partial_{\beta_s}
\Lambda_{\alpha_1 \cdots \alpha_{s-1}}.
\end{equation}
That is a direct link between these expressions and the traces of
non-local operators of the previous section. From this connection,
one can then construct a gauge invariant tensor $\cR_{\alpha_1
\cdots \alpha_{s};\beta_1 \cdots \beta_s}$ that is the proper
analogue of the Riemann tensor of a spin-2 field.

We can thus write in a more compact geometrical form the results
of the iterative procedure. The non-local field equations for odd
spin $s = 2n+1$ generalizing the Maxwell equations $\d^\mu
F_{\mu;\nu}=0$ are simply
\begin{equation}
\frac{1}{\square^{n}} \partial \cdot \cR^{[n]}_{;\mu_1 \cdots
\mu_{2n+1}} = 0 \ ,
\end{equation}
while the corresponding equations for even spin $s = 2n$ are
simply
\begin{equation}
\frac{1}{\square^{n-1}} \cR^{[n]}_{;\mu_1 \cdots \mu_{2n}} = 0 \ ,
\end{equation}
that reduce to $R_{;\mu\nu} = 0$ for spin 2.

The non-local geometric equations for higher-spin bosons can be
brought to the Fronsdal form using the traces $\Lambda^\prime$ of
the gauge parameters, and propagate the proper number of degrees
of freedom. At first sight, however, the resulting Fronsdal
equations present a subtlety \cite{Fra03}. The analysis of their
physical degrees of freedom normally rests on the choice of de
Donder gauge,
\begin{equation}
\cD \equiv \d \cdot \varphi - \half \, \d \varphi^{\prime} = 0 \ ,
\end{equation}
the higher-spin analog of the Lorentz gauge, that reduces the
Fronsdal operator to $\square \varphi$, but \emph{this is a proper
gauge only for doubly traceless fields}. The difficulty one faces
can be understood noting that, in order to recover the Fronsdal
equation eliminating the non-local terms, one uses the trace
$\Lambda'$ of the gauge parameter. The trace of the de Donder
gauge condition, proportional to the double trace $\varphi''$ of
the gauge field, is then in fact \emph{invariant} under residual
gauge transformations with a traceless parameter, so that the de
Donder gauge cannot be reached in general. However, it can be
modified, as in \cite{Fra03}, by the addition of terms containing
higher traces of the gauge field, and the resulting gauge fixed
equation then sets to the zero the double trace $\varphi''$ on
shell.

Following similar steps, one can introduce non-local equations for
\emph{fermionic} fields with unconstrained gauge fields and gauge
parameters \cite{Fra02a}. To this end, it is convenient to notice
that the fermionic operators for spin $s+1/2$ can be related to
the corresponding bosonic operators for spin $s$ according to
\begin{equation}
\cS_{s+1/2} - \half \frac{\d \sd}{\square} \scS_{s+1/2} = i
\frac{\sd}{\square} \cF_s(\psi) \ , \label{link}
\end{equation}
that generalize the obvious link between the Dirac and
Klein-Gordon operators. For instance, the Rarita-Schwinger
equation $\gamma^{\mu\nu\rho} \d_\nu \psi_\rho = 0$ implies that
\begin{equation}
\cS \equiv i(\sd \psi_\mu - \d_\mu \psis) = 0 \ ,
\end{equation}
while~(\ref{link}) implies that
\begin{equation}
S_\mu - \half \frac{\d_\mu \sd}{\square} \scS \ = \
\frac{i\sd}{\square} [\eta_{\mu\nu}- \d_\mu \d_\nu]\psi^\nu \ .
\end{equation}

Non-local fermionic kinetic operators $\cS^{(n)}$ can be defined
recursively as
\begin{equation}
\cS^{(n+1)} = \cS^{(n)} + \frac{1}{n(2n+1)} \frac{\d^2}{\square}
\cS^{(n)^\prime} - \frac{2}{2n+1} \frac{\d}{\square} \d \cdot
\cS^{(n)} \ ,
\end{equation}
with the understanding that, as in the bosonic case, the iteration
procedure stops when the gauge variation
\begin{equation}
\delta \cS^{(n)} = -2 i\, n\,
\frac{2n}{\square^{n-1}}\epss^{[n-1]}
\end{equation}
vanishes due to the impossibility of constructing the
corresponding higher trace of the gauge parameter. The key fact
shown in \cite{Fra02a,Fra03} is that, as in the bosonic case, the
Bianchi identities are similarly modified, according to
\begin{equation} \partial {\cal S}^{(n)} \ - \ \frac{1}{2n} \, \partial \
{\cal S}^{(n)\; '} \ - \ \frac{1}{2n}\, {\not {\! \partial}}
\scS^{(n)} \  = \ i \ \frac{\partial^{\; 2n}}{\square^{\; n-1}}
\psis^{[n]} \ , \end{equation}
and lack the anomalous terms when
$n$ is large enough to ensure that the field equations are fully
gauge invariant. Einstein-like operators and field equations can
then be defined following steps similar to those illustrated for
the bosonic case.

\section{Triplets and local compensator form}

\subsection{String field theory and BRST}

String Theory includes infinitely many higher-spin massive fields
with consistent mutual interactions, and it tensionless limit
$\alpha^\prime \rightarrow \infty$ lends itself naturally to
provide a closer view of higher-spin fields. Conversely, a better
grasp of higher-spin dynamics is likely to help forward our
current understanding of String Theory.

Let us recall some standard properties of the open bosonic string
oscillators. In the mostly plus convention for the metric, their
commutations relations read
\begin{equation}
[\alpha^\mu_k,\alpha^\nu_l] = k \delta_{k+l,0} \eta^{\mu\nu} \ ,
\end{equation}
and the corresponding Virasoro operators
\begin{equation}
L_k = \half \Sigma_{l=-\infty}^{+\infty} \alpha^\mu_{k-l} \alpha_{\mu l},
\end{equation}
where $\alpha_0^\mu = \sqrt{2\alpha^\prime} p^\mu$ and $p_\mu
-i\d_\mu$ satisfy the Virasoro algebra
\begin{equation}
[L_k,L_l] = (k-l)L_{k+l} + \frac{\cD}{12}m (m^2 - 1)\ ,
\end{equation}
where the central charge equals the space-time dimension $\cD$.

In order to study the tensionless limit, it is convenient to
rescale the Virasoro generators according to
\begin{equation}
L_k \rightarrow \frac{1}{\sqrt{2\alpha^\prime}} L_k,\qquad L_0 \rightarrow
\frac{1}{\alpha^\prime}L_0.
\end{equation}
Taking the limit $\alpha^\prime \rightarrow \infty$, one can then
define the reduced generators
\begin{equation}
l_0 = p^2, \quad l_m =  p\cdot\alpha_m \qquad (m\neq 0) \ ,
\end{equation}
that satisfy the simpler algebra
\begin{equation}
[l_k,l_l] = k \delta_{k+l,0} l_0 \ .
\end{equation}
Since this contracted algebra does not contain a central charge,
the resulting massless models are consistent in any space-time
dimension, in sharp contrast with what happens in String Theory
when $\alpha^\prime$ is finite. It is instructive to compare the
mechanism of mass generation at work in String Theory with the
Kaluza-Klein reduction, that as we have seen in previous sections
works for arbitrary dimensions. A closer look at the first few
mass levels shows that, \emph{as compared to the Kaluza-Klein
setting, the string spectrum lacks some auxiliary fields, and this
feature may be held responsible for the emergence of the critical
dimension!}

Following the general BRST method, let us introduce the ghost
modes $C_k$ of ghost number one and the corresponding antighosts
$B_k$ of ghost number minus one, with the usual anti-commutation
relations. The BRST operator \cite{Kat83,Wit85,Nev86}
\begin{equation}
\cQ = \sum_{-\infty}^{+ \infty} [C_{-k}L_k - \half
(k-l)\,:\,C_{-k}C_{-l}B_{k+l}:] - C_0
\end{equation}
determines the free string equation
\begin{equation}
\cQ | \Phi\rangle = 0 \ ,\label{eqstring}
\end{equation}
while the corresponding gauge transformation is
\begin{equation}
\delta | \Phi \rangle = \cQ | \Lambda \rangle\
 .\label{gaugestring}
\end{equation}
Rescaling the ghost variables according to
\begin{equation}
c_k = \sqrt{2\alpha^\prime} C_k \ , \quad b_k \frac{1}{2\alpha^\prime} B_k \ ,
\end{equation}
for $k \neq 0$ and as
\begin{equation}
c_0 = \alpha^\prime C_0, \quad b_0 = \frac{1}{\alpha^\prime} B_0
\end{equation}
for $k=0$ allows a non-singular $\alpha^\prime \rightarrow \infty$ limit
that defines the \emph{identically} nilpotent BRST charge
\begin{equation}
Q = \sum_{-\infty}^{+\infty} [c_{-k}l_k - \frac{k}{2} b_0
c_{-k}c_k]\ .
\end{equation}
Making the zero-mode structure manifest then gives
\begin{equation}
Q = c_0l_0 - b_0M + \tilde Q \ ,
\end{equation}
where $\tilde Q = \sum_{k\neq 0} c_{-k}l_k$ and $M=\half
\sum_{-\infty}^{+\infty} k c_{-k}c_k$, and the string field and
the gauge parameter can be decomposed as
\begin{eqnarray}
|\Phi\rangle &=& |\varphi_1\rangle + c_0 | \varphi_2 \rangle \ ,\\
|\Lambda\rangle &=& |\Lambda_1\rangle + c_0 | \Lambda_2 \rangle \
.
\end{eqnarray}

It should be appreciated that \emph{in this formulation no trace
constraint is imposed on the master gauge field $\varphi$ or on
the master gauge parameter $\Lambda$}. It is simple to confine the
attention to totally symmetric tensors, selecting states
$|\varphi_1\rangle$, $|\varphi_2\rangle$ and $|\Lambda\rangle$
that are built from a single string mode $\alpha_{-1}$,
\begin{eqnarray}
|\varphi_1\rangle &=& \sum_{s=0}^\infty \frac{1}{s!}\, \varphi_{\mu_1
\cdots
\mu_s}(x) \alpha^{\mu_1}_{-1} \cdots \alpha^{\mu_s}_{-1} |0\rangle \ ,\nonumber \\
&+&\sum_{s=2}^\infty \frac{1}{(s-2)!} D_{\mu_1 \cdots
\mu_{s-2}}(x) \alpha^{\mu_1}_{-1} \cdots \alpha^{\mu_{s-2}}_{-1} c_{-1}b_{-1}|0\rangle \ , \\
|\varphi_2\rangle &=& \sum_{s=1}^\infty \frac{-i}{(s-1)!} C_{\mu_1 \cdots
\mu_{s-1}}(x) \alpha^{\mu_1}_{-1} \cdots \alpha^{\mu_{s-1}}_{-1}b_{-1} |0\rangle \ ,\\
|\Lambda\rangle &=& \sum_{s=1}^\infty \frac{i}{(s-1)!}
\Lambda_{\mu_1 \cdots \mu_{s-1}}(x) \alpha^{\mu_1}_{-1} \cdots
\alpha^{\mu_{s-1}}_{-1} b_{-1}|0\rangle \ .
\end{eqnarray}
Restricting eqs.~(\ref{eqstring}) and (\ref{gaugestring}) to
states of this type, the $s$-th terms of the sums above yield the
\emph{triplet} equations \cite{Ben86,Hen87,Fra03}
\begin{eqnarray}
&&\square \varphi = \d C \ ,\nonumber\\
&&\d \cdot \varphi - \d D = C \ ,\\
&&\square D = \d \cdot C \ ,\nonumber
\end{eqnarray}
and the corresponding gauge transformations
\begin{eqnarray}
\delta \varphi &=& \d \Lambda \ ,\nonumber\\
\delta C &=& \square \Lambda \ ,\\
\delta D &=& \partial \cdot\Lambda \ ,\nonumber
\end{eqnarray}
where $\varphi$ is rank-$s$ tensor, $C$ is a rank-$(s-1)$ tensor
and $D$ is a rank-$(s-2)$ tensor. These field equations follow
from a corresponding truncation of the Lagrangian
\begin{equation}
\cL = \langle \Phi | Q | \Phi \rangle \ ,
\end{equation}
that in component notation reads
\begin{eqnarray}
\cL &=&  - \half (\d_\mu \varphi)^2 + s \d \cdot \varphi C +
s(s-1) \d \cdot C \, D +\frac{s(s-1)}{2} (\d_\mu D)^2 -
\frac{s}{2} C^2 \ ,
\end{eqnarray}
where the $D$ field, whose modes disappear on the mass shell, has
a peculiar negative kinetic term. Note that one can also eliminate
the auxiliary field $C$, thus arriving at the equivalent
formulation
\begin{eqnarray}
\cL &=& - \sum_{s=0}^{+\infty} \frac{1}{s!}\big[\half (\d_\mu
\varphi)^2 + \frac{s}{2} (\d \cdot \varphi)^2 + s(s-1) \d \cdot \d
\cdot \varphi \, D \nonumber \\ &+& s(s-1) (\d_\mu D)^2 +
\frac{s(s-1)(s-2)}{2} (\d\cdot D)^2\big] \ .
\end{eqnarray}
in terms of pairs $(\varphi,D)$ of symmetric tensors, more in the
spirit of \cite{Ben86}.

For a given value of $s$, this system propagates modes of spin
$s$, $s-2$, \dots, down to 0 or 1 according to whether $s$ is even
or odd. This can be simply foreseen from the light-cone
description of the string spectrum, since the corresponding
physical states are built from arbitrary powers of a single
light-cone oscillator $\alpha_1^i$, that taking out traces
produces precisely a nested chain of states with spins separated
by two units. From this reducible representation, as we shall see,
it is possible to deduce a set of equations for an irreducible
multiplet demanding that the trace of $\varphi$ be related to $D$.

If the auxiliary $C$ field is eliminated, the equations of motion for the
triplet take the form
\begin{eqnarray}
&& \cF = \d^2 (\varphi^\prime - 2D)\ ,\label{eqCD1}\\
&& \square D = \half \d \cdot \d \cdot \varphi - \half \d \d \cdot
D \ . \label{eqCD2}
\end{eqnarray}

\subsection{(A)dS extensions of the bosonic triplets}

The interaction between a spin 3/2 field and the gravitation field
is described essentially by a Rarita-Schwinger equation where
ordinary derivatives are replaced by Lorentz-covariant
derivatives. The gauge transformation of the Rarita-Schwinger
Lagrangian is then surprisingly proportional not to the Riemann
tensor, but to the Einstein tensor, and this variation is
precisely compensated in supergravity \cite{sugra} by the
supersymmetry variation of the Einstein Lagrangian. However, if
one tries to generalize this result to spin $s \geq 5/2$, the
miracle does not repeat and the gauge transformations of the field
equations of motion generate terms proportional to the Riemann
tensor, and similar problems are also met in the bosonic case.
This is the Aragone-Deser problem for higher spins \cite{Ara79}.

As was first noticed by Fradkin and Vasiliev \cite{Fra87}, with a
non-vanishing cosmological constant $\Lambda$ it is actually
possible to modify the spin $s \geq 5/2$ field equations
introducing additional terms that depend on negative powers of
$\Lambda$ and cancel the dangerous Riemann curvature terms. This
observation plays a crucial role in the Vasiliev equations
\cite{vasiliev}, discussed in the lectures by Vasiliev
\cite{vassolvay} and Sundell \cite{sundsolvay} at this Workshop.

For these reasons it is interesting to describe the (A)dS
extensions of the massless triplets that emerge from the bosonic
string in the tensionless limit and the corresponding deformations
of the compensator equations. Higher-spin gauge fields propagate
consistently and independently of one another in conformally flat
space-times, bypassing the Aragone-Deser inconsistencies that
would be introduced by a background Weyl tensor, and this
free-field formulation in an (A)dS background serves as a starting
point for exhibiting the unconstrained gauge symmetry of
\cite{Fra02a,Fra03}, as opposed to the constrained Fronsdal gauge
symmetry \cite{Fro78}, in the recent form of the Vasiliev
equations \cite{vasiliev} based on vector oscillators
\cite{sundsolvay}.

One can build the (A)dS symmetric triplets from a modified BRST
formalism \cite{Sag04}, but in the following we shall rather build
them directly deforming the flat triplets. The gauge
transformations of $\varphi$ and $D$ are naturally turned into
their curved-space counterparts,
\begin{eqnarray}
\delta \varphi &=& \nabla \Lambda \ , \label{eqgauge1} \\
\delta D &=& \nabla \cdot \Lambda \ ,\label{eqgauge2}
\end{eqnarray}
where the commutator of two covariant derivatives on a vector in
AdS is
\begin{equation}
[\nabla_\mu,\nabla_\nu]V_\rho = \frac{1}{L^2}(g_{\nu\rho}V_\mu -
g_{\mu\rho}V_\nu ) \ .\label{commBos}
\end{equation}
However, in order to maintain the definition of $C = \nabla  \cdot
\varphi - \nabla D$, one is led to deform its gauge variation,
turning it into
\begin{equation}
\delta C = \square \Lambda + \frac{(s-1)(3-s-\cD)}{L^2} \Lambda +
\frac{2}{L^2}g\Lambda^\prime \ , \label{eqgauge3}
\end{equation}
where $-1/L^2$ is the AdS cosmological constant and $g$ is the
background metric tensor. The corresponding de Sitter equations
could be obtained by the formal continuation of $L$ to imaginary
values, $L\rightarrow iL$.

These gauge transformations suffice to fix the other equations,
that read
\begin{eqnarray}
&&\square \varphi = \nabla C -\frac{1}{L^2}\left\{ -8gD + 2g\varphi^\prime
- [(2-s)(3-\cD - s) - s]\varphi
\right\} \ ,\label{eqAds}\\
&&C = \nabla \cdot \varphi - \nabla D,\\
&&\square D = \nabla \cdot C - \frac{1}{L^2}\left\{ -[s(\cD + s
-2) + 6]D + 4 \varphi^\prime + 2gD^\prime \right\} \ ,
\end{eqnarray}
and as in the previous section one can also eliminate $C$. To this
end, it is convenient to define the AdS Fronsdal operator, that
extends~(\ref{fronseq}), as
\begin{equation}
\cF \equiv \square \varphi - \nabla \nabla \cdot \varphi + \half
\{ \nabla,\nabla\}\varphi^\prime \ .
\end{equation}
The first equation of (\ref{eqAds}) then becomes
\begin{eqnarray}
\cF &=& \half \{ \nabla,\nabla\} \left( \varphi^\prime - 2D
\right)+ \frac{1}{L^2} \left\{ 8gD - 2g\varphi^\prime +
[(2-s)(3-\cD-s)-s]\varphi \right\} \ . \label{eqF}
\end{eqnarray}
In a similar fashion, after eliminating the auxiliary field $C$,
the AdS equation for $D$ becomes
\begin{eqnarray}
\square D + \half \nabla\nabla \cdot D &-&  \half \nabla\cdot \nabla \cdot
\varphi = - \frac{(s-2)(4-\cD -s)}{2L^2}D - \frac{1}{L^2} g D^\prime \nonumber\\
&+& \frac{1}{2L^2} \left\{ [s(\cD+s-2)+6]D-4\varphi^\prime -
2gD^\prime \right\} \ .
\end{eqnarray}
It is also convenient to define the modified Fronsdal operator
\begin{equation}
\cF_L = \cF - \frac{1}{L^2} \left\{ [(3-\cD-s)(2-s)-s]\varphi+2g
\varphi^\prime \right\}\ ,
\end{equation}
since in terms of $\cF_L$ eq.~(\ref{eqF}) becomes
\begin{equation}
\cF_L = \half \{ \nabla,\nabla \} (\varphi^\prime - 2D) +
\frac{8}{L^2}\,g\,D \ ,
\end{equation}
while the Bianchi identity becomes simply \beq \nabla \cdot \cF_L
- \half \nabla \cF^\prime_L = - \frac{3}{2} \nabla^3
\varphi^{\prime\prime} + \frac{2}{L^2}g\,\nabla
\varphi^{\prime\prime} \ . \eeq

\subsection{Compensator form of the bosonic equations}
\label{tripletscomp} In the previous sections we have displayed a
non-local geometric Lagrangian formulation for higher-spin bosons
and fermions. In this section we show how one can obtain very
simple \emph{local non-Lagrangian} descriptions that exhibit the
unconstrained gauge symmetry present in the non-local equations
and reduce to the Fronsdal form after a partial gauge fixing
\cite{Fra02a,Fra03,Sag04}.

The key observation is that the case of a single propagating
spin-$s$ field can be recovered from the equations
(\ref{eqCD1})-(\ref{eqCD2}) demanding that all lower-spin
excitations be pure gauge. To this end, it suffices to introduce a
spin $s-3$ \emph{compensator} $\alpha$ as
\begin{equation}
\varphi^\prime -2D = \d \alpha \ ,
\end{equation}
that by consistency transforms as
\begin{equation}
\delta \alpha = \Lambda' \ .
\end{equation}
Eq.~(\ref{eqCD1}) then becomes
\begin{equation}
\cF = 3 \d^3 \alpha \ ,
\end{equation}
while (\ref{eqCD2}) becomes
\begin{equation}
\cF^\prime -  \d^2 \varphi^{\prime\prime} = 3\;  \square \d \alpha
\ + \ 2 \; \d^2 \d \cdot\alpha \ .
\end{equation}
Combining them leads to
\begin{equation}
\d^2 \varphi^{\prime\prime} \ = \
\d^2(4\d\cdot\alpha+\d\alpha^\prime)\ ,
\end{equation}
and the conclusion is then that the triplet equations imply a pair
of \emph{local} equations for a single massless spin-$s$ gauge
field $\varphi$ and a single spin-$(s-3)$ compensator $\alpha$.
Summarizing, the local compensator equations and the corresponding
gauge transformations are
\begin{equation}\begin{array}{ll}
\cF = 3 \d^3 \alpha
 ,& \varphi^{\prime\prime} = 4\d \cdot
\alpha+\d
\alpha^\prime
\ ,\\
\delta \varphi = \d \Lambda,& \delta \alpha = \Lambda^\prime \ ,
\end{array} \label{eqcmp}
\end{equation}
and clearly reduce to the standard Fronsdal form after a partial gauge
fixing using the trace $\Lambda^\prime$ of the gauge parameter . These
equations can be regarded as the local analogs of the non-local geometric
equations, but it should be stressed that \emph{they are not Lagrangian
equations}. This can be seen either directly, as in \cite{Fra03,Sag04}, or
via the corresponding BRST operator, that is not hermitian, as pertains to
a reduced system that is not described by a Lagrangian \cite{Bar04}.
Nonetheless, the two equations (\ref{eqcmp}) form a consistent system, and
the first can be turned into the second using the Bianchi identity.

One can also obtain the (A)dS extension of the spin-$s$ compensator
equations~(\ref{eqcmp}). The natural starting point are the (A)dS gauge
transformations for the fields $\varphi$ and $\alpha$
\begin{equation}
\delta \varphi = \nabla \Lambda \ ,\qquad \delta \alpha \Lambda^\prime \ .
\end{equation}
One can then proceed in various ways to obtain the compensator equations
\begin{eqnarray}
\cF &=& 3 \nabla^3\alpha + \frac{1}{L^2} \left\{-2g\varphi^\prime
+ [(2-s)(3-\cD-s)-s]\varphi \right\}- \frac{4}{L^2}g\nabla\alpha \
\nonumber \\
&& \varphi'' \ = \ 4\d\cdot\alpha+\d\alpha^\prime \
,\label{eqFAds}
\end{eqnarray}
that, of course, again do \emph{not} follow from a Lagrangian.
However, lagrangian equations can be obtained, both in flat space
and in an (A)dS background, from a BRST construction based on a
wider set of constraints first obtained by Pashnev and Tsulaia
\cite{Pas97,Sag04}. It is instructive to illustrate these results
for spin 3 bosons.

In addition to the triplet fields $\varphi$, $C$ and $D$ and the
compensator $\alpha$, this formulation uses the additional spin-1
fields $\varphi^{(1)}$ and $F$ and spin-0 fields $C^{(1)}$ and
$E$, together with a new spin-1 gauge parameter $\mu$ and a new
spin-0 gauge parameter $\Lambda^{(1)}$. The BRST analysis
generates the gauge transformations
\begin{equation}
\begin{array}{ll} \delta \varphi = \d \Lambda + \eta \; \mu \ ,& \delta \alpha = \Lambda^\prime - \sqrt{2\cD}
\Lambda^{(1)} \ ,\\ \delta \varphi^{(1)} = \d
\Lambda^{(1)}+\sqrt{\frac{\cD}{2}}\mu \ , & \delta C = \square \Lambda \ ,\\
\delta D = \d \cdot \Lambda + \mu \ , &\delta C^{(1)} = \square
\Lambda^{(1)} \ ,\\
\delta E = \d \cdot \mu \ , &\delta F = \square \mu\ ,
\end{array} \label{spin3offshell}
\end{equation}
and the corresponding field equations
\begin{equation}
\begin{array}{ll} \square \varphi = \d C +\eta F \ ,& \square \alpha C^\prime - \sqrt{2\cD} C^{(1)}\ ,\\
\d\cdot\varphi-\d D-\eta E = C \ ,& \square \varphi^{(1)} = \d
C^{(1)}
+\sqrt\frac{\cD}{2}F \ ,\\
\square D = \d \cdot C + F \ ,& \square E = \d \cdot F \ ,\\
\d \alpha = \varphi^{\prime} - 2D-\sqrt{2\cD}\varphi^{(1)} \ ,& \d
\cdot \varphi^{(1)} - \sqrt{2\cD}E = C^{(1)} \ .
\end{array}
\end{equation}
Making use of the gauge parameters $\mu$ and $\Lambda^{(1)}$ one
can set $\varphi^{(1)}=0$ and  $C^{(1)} = 0$, while the other
additional fields are set to zero by the field equations.
Therefore, one can indeed recover the non-Lagrangian compensator
equations gauge fixing this Lagrangian system. A similar, if more
complicated analysis, goes through for higher spins, where this
formulation requires ${\cal O}(s)$ fields.

The logic behind these equations can be captured rather simply
taking a closer look at the gauge transformations
(\ref{spin3offshell}). One is in fact gauging away $\varphi'$,
modifying the gauge transformation of $\varphi$ by the $\mu$ term.
This introduces a corresponding modification in the $\varphi$
equation, that carries through by integrability to the $C$
equation, and so on.

\subsection{Fermionic triplets}

We can now turn to the fermionic triplets proposed in \cite{Fra03}
as a natural guess for the field equations of symmetric
spinor-tensors arising in the tensionless limit of superstring
theories. In fact, the GSO projection limits their direct
occurrence to type-0 theories \cite{type0}, but slightly more
complicated spinor-tensors of this type, but with mixed symmetry,
are present in all superstring spectra, and can be discussed along
similar lines \cite{Sag04}.

The counterparts of the bosonic triplet equations and gauge
transformations are
\begin{equation}
\begin{array}{ll} \sd \psi = \d \chi \ ,& \delta \psi = \d \eps \ ,\\
\d \cdot \psi -\d \lambda = \sd \chi \ ,\qquad & \delta \lambda = \d \cdot \eps \ ,\\
\sd \lambda = \d \cdot \chi \ , & \delta \chi = \sd \eps \ .
\end{array}
\end{equation}
It can be shown that this type of system propagates spin-$(s+1/2)$
modes and \emph{all} lower half-integer spins. One can now
introduce a spin-$(s-2)$ compensator $\xi$ proceeding in a way
similar to what we have seen for the bosonic case, and the end
result is a simple non-Lagrangian formulation for a single
spin-$s$ field,
\begin{equation}
\begin{array}{ll}
\cS = -2 \; i\; \d^2 \xi \ ,& \delta \psi = \d \eps \ ,\\
\psis^\prime = 2 \; \d \cdot \xi + \d \xi^\prime + \sd \xis \ ,
\qquad& \delta \xi = \epss \ .
\end{array} \label{eq45}
\end{equation}
These equations turn into one another using the Bianchi identity,
and can be extended to (A)dS background, as in \cite{Sag04}.
However, a difficulty presents itself when attempting to extend
the fermionic triplets to off-shell systems in (A)dS, since the
BRST analysis shows that the operator extension does not define a
closed algebra.

\subsection{Fermionic compensators}

One can also extend nicely the fermionic compensator equations to
an (A)dS background. The gauge transformation for a spin-$(s+
1/2)$ fermion is deformed in a way that can be anticipated from
supergravity and becomes in this case
\begin{equation}
\delta \psi = \nabla \eps + \frac{1}{2L}\gamma \eps \ ,
\end{equation}
where $L$ determines again the (A)dS curvature and $\nabla$
denotes an (A)dS covariant derivative. The commutator of two of
these derivatives on a spin-$1/2$ field $\eta$ reads
\begin{equation}
[\nabla_\mu,\nabla_\nu]\eta = - \frac{1}{2L^2}\;
\gamma_{\mu\nu}\eta\ , \label{commFerm}
\end{equation}
and using eqs~(\ref{commBos})-(\ref{commFerm}) one can show that
the compensator equations for a spin-$s$ fermion ($s = n+1/2$) in
an (A)dS background are
\begin{eqnarray}
(\nablas \psi - \nabla \psis) &+ & \frac{1}{2L}[\cD +
2(n-2)]\psi+\frac{1}{2L}\gamma \psis \nonumber\\
&=& - \{\nabla,\nabla\}\xi + \frac{1}{L} \gamma \nabla \xi +
\frac{3}{2L^2}g\xi \ , \label{eq46}\\
\psis^\prime &=& 2 \nabla \cdot \xi + \nablas \xis + \nabla
\xi^\prime + \frac{1}{2L} [ \cD + 2(n-2)]\xis - \frac{1}{2L}\gamma
\xi^\prime . \nonumber
\end{eqnarray}
These equations are invariant under
\begin{eqnarray}
\delta \psi &=& \nabla \eps \ , \\
\delta \xi &=& \epss \ ,
\end{eqnarray}
with an unconstrained parameter $\eps$. Eqs. (\ref{eq46}) are
again a pair of non-lagrangian equations, like their flat space
counterparts (\ref{eq45}).

As in the flat case, eqs~(\ref{eq46}) are nicely consistent, as
can be shown making use of the (A)dS deformed Bianchi
identity~(\ref{bianchihalf})
\begin{eqnarray}
\nabla \cdot \cS - \half \nabla \cS^\prime - \half \nablas \scS &=&
\frac{i}{4L}\gamma \cS^\prime + \frac{i}{4L}[(\cD - 2)+ 2(n-1)] \scS \\
&+& \frac{i}{2} \left[ \{\nabla,\nabla\} - \frac{1}{L}\gamma
\nabla - \frac{3}{2L^2} \right] \psis^\prime \ ,
\end{eqnarray}
where the Fang-Fronsdal operator $\cS$ is also deformed and
becomes
\begin{equation}
\cS = i (\nablas \psi - \nabla \psis) + \frac{i}{2L}[\cD +
2(n-2)]\psi + \frac{i}{2L} \gamma \psis \ .
\end{equation}

\section*{Acknowledgments}

N.B. and G.C. would like to thank Sophie de Buyl and Christiane
Schomblond for their enlightening remarks and their careful
reading of the manuscript. A.S. would like to thank the Organizers
for their kind invitation to present these lectures and the
participants for a very stimulating atmosphere. The work of N.B.
and G.C. was supported in part by the ``P\^{o}le d'Attraction
Interuniversitaire'' (Belgium), by IISN-Belgium (convention
4.4505.86), by Proyectos FONDECYT 1970151 and 7960001 (Chile) and
"Le Fonds National de la Recherche Scientifique (Belgium)". The
work of A.S. was supported in part by INFN, by the EU contracts
HPRN-CT-2000-00122 and HPRN-CT-2000-00148, by the MIUR-COFIN
contract 2003-023852, by the INTAS contract 03-51-6346 and by the
NATO grant PST.CLG.978785.

\bibliography{}

\begin{thebibliography}{2}

\bibitem{Fro78} C.~Fronsdal,
Phys.\ Rev.\ D {\bf 18}, 3624 (1978).

\bibitem{Fro79} J.~Fang and C.~Fronsdal,
Phys.\ Rev.\ D {\bf 18}, 3630 (1978).

\bibitem{Fra02a} D.~Francia and A.~Sagnotti,
Phys.\ Lett.\ B {\bf 543}, 303 (2002) [arXiv:hep-th/0207002].

\bibitem{Fra03} D.~Francia and A.~Sagnotti,
Class.\ Quant.\ Grav.\ {\bf 20}, S473 (2003) [arXiv:hep-th/0212185].

\bibitem{Sag04} A.~Sagnotti and M.~Tsulaia,
Nucl.\ Phys.\ B {\bf 682}, 83 (2004) [arXiv:hep-th/0311257].

\bibitem{Fie39} M.~Fierz and W.~Pauli,
Proc.\ Roy.\ Soc.\ Lond.\ A {\bf 173}, 211 (1939).

\bibitem{Wig39} E.~P.~Wigner,
Annals Math.\  {\bf 40}, 149 (1939) [Nucl.\ Phys.\ Proc.\ Suppl.\  {\bf
6}, 9 (1989)].

\bibitem{Bar48} V.~Bargmann and E.~P.~Wigner,
Proc. Natl. Acad. Sci. USA \textbf{34} (1948) 211.

\bibitem{Sin74} L.~P.~S.~Singh and C.~R.~Hagen,
Phys.\ Rev.\ D {\bf 9}, 898 (1974);
L.~P.~S.~Singh and C.~R.~Hagen,
Phys.\ Rev.\ D {\bf 9}, 910 (1974).

\bibitem{sugra} D.~Z.~Freedman, P.~van Nieuwenhuizen and S.~Ferrara,
Phys.\ Rev.\ D {\bf 13} (1976) 3214;
S.~Deser and B.~Zumino,
Phys.\ Lett.\ B {\bf 62} (1976) 335.

\bibitem{Bou02} P.~de Medeiros and C.~Hull,
Commun.\ Math.\ Phys.\  {\bf 235} (2003) 255 [arXiv:hep-th/0208155];
X.~Bekaert and N.~Boulanger,
Commun.\ Math.\ Phys.\  {\bf 245}, 27 (2004) [arXiv:hep-th/0208058];
X.~Bekaert and N.~Boulanger,
Phys.\ Lett.\ B {\bf 561}, 183 (2003) [arXiv:hep-th/0301243].

\bibitem{chrisolvay} C.~M.~Hull, to appear in the Proceedings of the First Solvay Workshop
on Higher Spin Gauge Theories (Brussels, May 2004).

\bibitem{massiveKK}
T.~R.~Govindarajan, S.~D.~Rindani and M.~Sivakumar,
Phys.\ Rev.\ D {\bf 32} (1985) 454;
S.~D.~Rindani and M.~Sivakumar,
Phys.\ Rev.\ D {\bf 32} (1985) 3238;
C.~Aragone, S.~Deser and Z.~Yang,
Annals Phys.\  {\bf 179}, 76 (1987).

\bibitem{deW80} B.~de Wit and D.~Z.~Freedman,
Phys.\ Rev.\ D {\bf 21}, 358 (1980);
T.~Damour and S.~Deser,
Annales Poincare Phys.\ Theor.\  {\bf 47} (1987) 277.

\bibitem{RS} W.~Rarita and J.~S.~Schwinger,
Phys.\ Rev.\  {\bf 60}, 61 (1941).

\bibitem{supergravity} D.~Z.~Freedman, P.~van Nieuwenhuizen and S.~Ferrara,
Phys.\ Rev.\ D {\bf 13}, 3214 (1976);
S.~Deser and B.~Zumino,
Phys.\ Lett.\ B {\bf 62}, 335 (1976).

\bibitem{Kat83} M.~Kato and K.~Ogawa,
Nucl.\ Phys.\ B {\bf 212}, 443 (1983).

\bibitem{Wit85} E.~Witten,
Nucl.\ Phys.\ B {\bf 268} (1986) 253.

\bibitem{Nev86} A.~Neveu and P.~C.~West,
Nucl.\ Phys.\ B {\bf 268} (1986) 125.

\bibitem{Ben86} A.~K.~H.~Bengtsson,
Phys.\ Lett.\ B {\bf 182} (1986) 321.

\bibitem{Hen87} M.~Henneaux and C.~Teitelboim,
``First And Second Quantized Point Particles Of Any Spin,'' Chap 9, pp.
113, ``Quantum Mechanics of Fundamental Systems, 2,'' eds. C. Teitelboim
and J. Zanelli (Plenum Press, New York, 1988).


\bibitem{Ara79} C.~Aragone and S.~Deser,
Phys.\ Lett.\ B {\bf 86}, 161 (1979).

\bibitem{Fra87} E.~S.~Fradkin and M.~A.~Vasiliev,
Phys.\ Lett.\ B {\bf 189}, 89 (1987);
E.~S.~Fradkin and M.~A.~Vasiliev,
Nucl.\ Phys.\ B {\bf 291}, 141 (1987).

\bibitem{vasiliev} M.~A.~Vasiliev,
Phys.\ Lett.\ B {\bf 243}, 378 (1990),
Class.\ Quant.\ Grav.\  {\bf 8}, 1387 (1991),
Phys.\ Lett.\ B {\bf 257}, 111 (1991);
M.~A.~Vasiliev,
Phys.\ Lett.\ B {\bf 285}, 225 (1992),
Phys.\ Lett.\ B {\bf 567}, 139 (2003) [arXiv:hep-th/0304049].
For reviews see: M.~A.~Vasiliev,
Int.\ J.\ Mod.\ Phys.\ D {\bf 5}, 763 (1996) [arXiv:hep-th/9611024],
arXiv:hep-th/9910096,
arXiv:hep-th/0104246.

\bibitem{vassolvay}M.~A.~Vasiliev, to appear in the Proceedings of the First Solvay Workshop
on Higher Spin Gauge Theories (Brussels, May 2004).

\bibitem{sundsolvay}P.~Sundell, to appear in Proceedings of the First Solvay Workshop
on Higher Spin Gauge Theories (Brussels, May 2004); A.~Sagnotti, E.~Sezgin
and P.~Sundell, to appear.


\bibitem{Bar04} G.~Barnich, M.~Grigoriev, A.~Semikhatov and I.~Tipunin,
arXiv:hep-th/0406192.


\bibitem{Pas97} A.~Pashnev and M.~M.~Tsulaia,
Mod.\ Phys.\ Lett.\ A {\bf 12}, 861 (1997)
[arXiv:hep-th/9703010].


\bibitem{type0} L.~J.~Dixon and J.~A.~Harvey,
Nucl.\ Phys.\ B {\bf 274}, 93 (1986);
N.~Seiberg and E.~Witten,
Nucl.\ Phys.\ B {\bf 276}, 272 (1986);
M.~Bianchi and A.~Sagnotti,
Phys.\ Lett.\ B {\bf 247}, 517 (1990);
A.~Sagnotti,
arXiv:hep-th/9509080,
Nucl.\ Phys.\ Proc.\ Suppl.\  {\bf 56B}, 332 (1997)
[arXiv:hep-th/9702093];
C.~Angelantonj,
Phys.\ Lett.\ B {\bf 444}, 309 (1998) [arXiv:hep-th/9810214].
R.~Blumenhagen, A.~Font and D.~Lust,
Nucl.\ Phys.\ B {\bf 558}, 159 (1999) [arXiv:hep-th/9904069].
For a review see: C.~Angelantonj and A.~Sagnotti,
Phys.\ Rept.\  {\bf 371}, 1 (2002) [Erratum-ibid.\  {\bf 376}, 339 (2003)]
[arXiv:hep-th/0204089].

\end{thebibliography}

\end{document}